
\font\mybb=msbm10 at 12pt
\def\bbvv#1{\hbox{\mybb#1}}

\def\R {\bbvv{R}}
\tolerance=10000
\input phyzzx

\def \aa {\alpha}
\def \bb {\beta}
\def \gg {\gamma}
\def \dd {\delta}
\def \ee {\epsilon}
\def \ffi {\phi}
\def \ff {\phi}
\def \kk {\kappa}
\def \ll {\lambda}
\def \mm {\mu}

\def \ss {\sigma}

\def \th {\theta}
\def \cc {\chi}
\def \ww{\omega}

 \def \ggg {\Gamma}
\def \ddd {\Delta}
\def \eee {\varepsilon}
\def \fff {\Phi}
\def \lll {\wedge}

\def \uuu {\Upsilon}

\def \ti {\tilde}
\def \ba {\bar}
\def\bbb{{\bar \beta}}
\def\aab{{\bar \alpha}}
\def\bbt{{\tilde \beta}}
\def\aat{{\tilde \alpha}}
\def\ab{{\bar \alpha}}

\def \2 {{1 \over 2}}
\def \3 {{1 \over 3}}
\def \4 {{1 \over 4}}
\def \5 {{1 \over 5}}
\def \6 {{1 \over 6}}
\def \7 {{1 \over 7}}
\def \8 {{1 \over 8}}
\def \9 {{1 \over 9}}
\def \0 { \infty}

\def\++ {{(+)}}
\def \- {{(-)}}
\def\+-{{(\pm)}}

\def\ek {\eqn\abc$$}
\def \ap {{\alpha '}}
\def   \gab {g_{\alpha \bar \beta} }

\def \pa {\partial}


 \def\unit{\hbox to 3.3pt{\hskip1.3pt \vrule height 7pt width .4pt \hskip.7pt
\vrule height 7.85pt width .4pt \kern-2.4pt
\hrulefill \kern-3pt
\raise 4pt\hbox{\char'40}}}
\def\II{{\unit}}

\def\gij{g_{ij}}
\def\bij{b_{ij}}
\def\fii{\phi^i}

\def\nup#1({Nucl.\ Phys.\  {\bf B#1}\ (}

\def\tr{{\rm tr}}

\REF\MK{D. Kutasov and E. Martinec, Nucl. Phys. {\bf B477} (1996) 652;
hep-th/9602049.}
\REF\MKO{D. Kutasov and E. Martinec and M. O'Loughlin, Nucl. Phys. {\bf B477}
(1996) 675;
hep-th/9603116.}
\REF\MKA{D. Kutasov and E. Martinec,  hep-th/9612102.}
\REF\Mucsb{E. Martinec,  hep-th/9608017.}
\REF\Mmat{T. Banks, W. Fischler, S. Shenker and L. Susskind, hep-th/9610043.}
\REF\Oo{H. Ooguri and C. Vafa, Mod. Phys. Lett. {\bf A5} (1990) 1389; Nucl.
Phys. {\bf 361} (1991)
469.}
\REF\hto{ C.M. Hull,  Phys. Lett. {\bf  B387} (1996) 497.}
\REF\Oot{H. Ooguri and C. Vafa,   Nucl. Phys. {\bf 367} (1991)
83.}
\REF\HW{C.M. Hull and E. Witten, Phys. Lett. {\bf 160B} (1985)  398.}
\REF\Hsig{ C.M. Hull, Nucl. Phys. {\bf B267} (1986)  266.}
\REF\Hsiga{ C.M. Hull,  Phys. Lett. {\bf 178B} (1986) 357.}
\REF\Hsigb{ C.M. Hull, in the Proceedings of the First Torino Meeting on
Superunification and Extra Dimensions,  edited by R. D'Auria and P. Fr\' e,
(World Scientific, Singapore, 1986).}
\REF\Van{ C.M. Hull, in Super Field Theories  (Plenum, New York, 1988), edited
by H. Lee and G. Kunstatter.}
\REF\klein{J. Barrett, G.W. Gibbons, M.J. Perry, C.N. Pope and P. Ruback, Int.
J. Mod. Phys.
{\bf A9} (1994) 1457.}
\REF\hit{N.J. Hitchin, {\it Hypersymplectic Quotients}, in Acta Academie
Scientiarum Taurinensis,
Supplemento al Numero 124 degli Atti della Accademia delle Scienze di Torino
Classe di Scienze
Fisiche, Matematiche e Naturali (1990).}
\REF\Don{S. Donaldson, Proc. Lond. Math. Soc., {\bf 50} (1985) 1; V. Nair and
J. Shiff, Nucl. Phys. {\bf B371} (1992) 329; Phys.
Lett. {\bf 246B} (1990) 423.}
\REF\Bch{R. Bott and S-S. Chern, Acta Math {\bf 114} (1965) 71.}
\REF\Bcha{R. Bott and S-S. Chern, Essays on Topology and Related Topics,
Springer-Verlag (1970)
48.}
\REF\GHR{S.J. Gates Jr., C.M. Hull and M. Ro\v cek, Nucl. Phys. {\bf B248}
(1984)  157.}
 \REF\HTa{C.M. Hull and P.K. Townsend,   Phys. Lett. {\bf 178B} (1986) 187.}
\REF\HPf{P.S. Howe and G. Papadopoulos, Class. Quantum Grav. {\bf 5} (1988)
1647.}
 \REF\Kou{E. Kiritsis, C. Kounnas and D. Lust, Int. J. Mod. Phys. {\bf A9}
(1994) 1361.}
\REF\prep{C.M. Hull, in preparation.}
\REF\Ya{K. Yano and M. Kon, {\it Structures on Manifolds}, World Scientific,
SIngapore, 1984.}
\REF\mohab{M. Abou Zeid and C.M. Hull, hep-th/9612208.}
 \REF\HP{P.S. Howe and G. Papadopoulos, Nucl. Phys. {\bf B289} (1987) 264.}


\Pubnum{ \vbox{ \hbox {QMW-97-2}  \hbox {NI97005}\hbox{hep-th/9702067}} }
\pubtype{}
\date{February, 1997}

\titlepage

\title {\bf  Actions For $(2,1)$ Sigma-Models and Strings}

\author{C.M. Hull}
\address{Physics Department,
Queen Mary and Westfield College,
\break
Mile End Road, London E1 4NS, U.K.}
\andaddress{Isaac Newton Institute, 20 Clarkson Road,
\break
Cambridge CB3 0EH, U.K.}
\vskip 0.5cm

\abstract {
Effective actions are derived for (2,0) and (2,1) superstrings by studying the
corresponding sigma-models.
 The geometry is a
generalisation of Kahler geometry involving torsion  and the field equations
imply that the
curvature with torsion is self-dual in four dimensions, or has $SU(n,m)$
holonomy in other
dimensions. The Yang-Mills fields are self-dual in four dimensions and satisfy
a form of
the Uhlenbeck-Yau equation  in higher dimensions.
In four dimensions with Euclidean signature, there is a hyperkahler structure
and the
sigma-model has (4,1) supersymmetry, while for signature (2,2) there is a
hypersymplectic structure
consisting of a complex
structure  squaring to $-{\II}$ and  two
\lq real structures' squaring to $ {\II}$. The theory is
invariant under  a  twisted
 form of the (4,1) superconformal algebra which includes an $SL(2,\R)$
Kac-Moody
algebra instead of an $SU(2)$ Kac-Moody algebra. Kahler and related geometries
 are generalised to
ones involving real structures.
}

\endpage
\pagenumber=1

\chapter {Introduction}

Martinec and Kutasov [\MK-\MKA] have argued that different vacua of the
superstring with (2,1)
world-sheet supersymmetry correspond to the 11-dimensional membrane, the type
IIB string,  the
heterotic and the type I strings. This suggests that the (2,1) string could be
useful in the search
for  the  degrees of freedom appropriate for the description of the fundamental
theory underpinning
M-theory and superstring theory. (Another proposal is given by the matrix model
[\Mmat].)  It was shown in
[\Oo] that the usual  string with (2,2) supersymmetry is a theory of self-dual
gravity in a
four-dimensional space-time with signature (2,2) governed by the  Plebanski
action, while in [\hto] it
was argued that the (2,2) supersymmetric string based on twisted chiral
multiplets is a theory of
self-dual gravity with torsion, which turns out to be a free theory. The (2,1)
and (2,0) strings
[\Oot] are again formulated in a four-dimensional space-time with signature
(2,2), but now a null
reduction must be imposed to obtain a space with signature (2,1) (which
corresponds to a membrane
world-volume [\MK]) or a space with signature (1,1) (which corresponds to a
string world-sheet
[\MK]). The (2,2)-dimensional theory before null-reduction is a theory of
gravity with torsion
coupled to Yang-Mills gauge fields. The purpose of this paper is to investigate
further  the
  target space geometry of (2,0) and (2,1) strings and sigma-models.

We give an effective action for the gravitational and anti-symmetric tensor
degrees of freedom of
the (2,1) string which was obtained independently by Martinec and Kutasov
[\MKA], who also proposed
a generalisation to include the Yang-Mills fields and  checked that this
agrees with the S-matrix
of (2,1) strings to quartic order in the fields.  A sigma-model derivation of
this action is given
and generalised to   give an effective  action whose variation gives the
conditions found in
[\Hsig-\Van] for conformal invariance of general (2,0) and (2,1) sigma-models.
The geometry is a
generalisation of Kahler geometry with torsion [\HW] and the field equations
imply that the
curvature with torsion is self-dual in four dimensions, or has $SU(n,m)$
holonomy in other
dimensions. In four dimensions with Euclidean signature, there is a hyperkahler
structure and the
sigma-model has (4,1) supersymmetry, while for signature (2,2) there is a
hypersymplectic structure
[\klein,\hit] -- instead of three complex structures squaring to $-{\II}$,
there is a complex
structure and  two
\lq real structures' or \lq locally product structures' squaring to $+{\II}$ --
and the model is
invariant under  a twisted  form of the (4,1) superconformal algebra which
includes an $SL(2,\R)$ Kac-Moody
algebra instead of an $SU(2)$ Kac-Moody algebra. Kahler and related geometries
 are generalised to
ones involving real structures. The Yang-Mills fields are self-dual in four
dimensions and satisfy
the Uhlenbeck-Yau equation
$g^{\aa{\bar \beta}}F_{\aa{\bar \beta}}=0$ in higher dimensions, but where the
metric  $g_{\aa{\bar
\beta}}$ involves  corrections dependent on the gauge-fields. The action is
related to that of
[\Don], and involves the Bott-Chern form [\Bch,\Bcha].

The results regarding the amount of supersymmetry can be summarised in terms of
the holonomy of a
 certain connection, which will have torsion in general.
For Euclidean signature, the holonomy of a general $D$-dimensional manifold $M$
is
$O(D) $, and a (1,1) supersymmetric sigma-model can  be defined on $M$.
If $D=2n$ and the holonomy ${\cal H} $ is in $U(n)$, there is a covariantly
constant
complex structure $J$, $J^2=-\II$, and the (1,1) model in fact has (2,1)
supersymmetry.
If ${\cal H} $ is in $SU(n)$, then there is a covariantly constant spinor and
so such a space preserves some space-time supersymmetry, and the space is a
solution of   string theory (or related to one by a certain deformation). In
the case of compact $M$ with vanishing torsion, these are the Calabi-Yau
spaces.
If $n=2m$ and ${\cal H} \subseteq USp(m)$,  (where $USp(m)$ is compact, with
the convention that
$USp(1)=SU(2)$)
then there is a covariantly constant
hyperkahler structure consisting of three complex structures
$ I,J,K $   satisfying the quaternion
algebra
$$\eqalign{&I^2=J^2=K^2=- {\II}, \cr &
IJ=-JI=K, JK=-KJ=I,  KI=-IK=J
\cr}
\eqn\quat$$
and the (1,1) model has (4,1) supersymmetry.

These results [\Van] for Euclidean signatures are well-known, but they can be
generalised to other
signatures. For signature $(2n_1,2n_2)$, if ${\cal H} $ is in $U(n_1,n_2)$
then there is again a
complex structure and (2,1) supersymmetry, and the generalised Calabi-Yau
condition is
${\cal H} \subseteq SU(n_1,n_2)$. Consider now the case of signature $(d,d)$,
for which the name
Kleinian geometry was suggested in [\klein]; the case $d=2$ is relevant for
$(2,p)$ strings. In
general the holonomy is in $O(d,d)$, but if $d=2n$ and
${\cal H} \subseteq U(n ,n )$ there is a complex structure $J$ leading to (2,1)
supersymmetry and the
Calabi-Yau-type condition is
${\cal H} \subseteq SU(n ,n )$. If on the other hand ${\cal H} \subseteq
GL(n,\R)$, then  there is a
real structure $S$ satisfying $S^2=\II$ and there is  an extra supersymmetry,
but the right-handed
superalgebra is of the form
$$\{ Q^A,Q^B \}=\eta ^{AB} P
\eqn\erte$$ where $A=1,2$ and $\eta ^{AB}= diag (1,-1)$ and $P$ is the
right-moving momentum.
If there is no torsion, then the metric is given in terms of a scalar
 potential analogous to the Kahler potential,
while if there is torsion, both the metric and torsion are given in terms of a
vector potential,
analogous to the one in [\HW].
The condition
${\cal H}
\subseteq SL(n,\R)$ is the
analogue of the Calabi-Yau condition; it implies Ricci-flatness if there is no
torsion,
or the   generalisation of this that corresponds to the string field equations
if there is torsion.
Finally, if
$n=2m$ and ${\cal H} \subseteq Sp(m,\R)$ (where  $Sp(m,\R)$ is non-compact,
with $Sp(1)=SL(2,\R)$), then
there are  three covariantly constant tensors
$ J,S,T$   satisfying the pseudoquaternion algebra [\klein,\hit]
$$
\eqalign{& J^2 =- {\II}, S^2=T^2={\II}\cr &
ST=-TS=-J, TJ=-JT=S,  JS=-SJ=T
\cr}
\eqn\slquat$$
$J$ is a complex structure and $S,T$ are real structures and the sigma-model
again has a twisted $(4,1)$ superconformal symmetry. The right-handed
superalgebra is again of the form
\erte,
where now $A=1,2,3,4$ and $\eta ^{AB}= diag (1,1,-1,-1)$.
As $Sp(m,\R)$ is a subgroup of  both $SU(m,m)$ and $SL(2m,\R)$, such spaces
give string solutions.
 For $m=1$, $SU(1,1)=SL(2,\R)=Sp(1,\R)$ and spaces with this holonomy have
self-dual curvature (with
torsion).

\chapter{The (2,1) Supersymmetric Sigma-Model}

The (1,1) supersymmetric  sigma-model   with metric $\gij$ and anti-symmetric
tensor $\bij$
has the (1,1) superspace action [\GHR]
$$S= \int d^2x d^2 \theta \, [\gij + \bij] D_+ \phi ^i D_-\phi ^j
\eqn\acton$$
It will be conformally invariant at one-loop if there is a function
 $\Phi$ such that
$$
R^{(+)} _{ij}-2 \nabla_{(i} \nabla _{j)} \Phi- 2H_{ij}{}^k \nabla _k\Phi
=0\eqn\conf$$
where $
R^{(+)} _{ij}$ is the Ricci tensor for a connection with torsion.
We define the connections with torsion
$$ \Gamma ^{(\pm) i}_{jk}=
\left\{ {i \atop jk} \right\}
\pm H^i_{jk}\eqn\con$$
where  $\left\{ {i \atop jk} \right\}$ is the Christoffel connection and the
torsion tensor is
$$
H_{ijk}= {3 \over 2}\partial_{[i}b_{jk]}
\eqn\abc$$
The
curvature and Ricci tensors with torsion are
$$ R^{(\pm) k} {}_{lij}= \partial_i \Gamma^{(\pm) k}_{jl}-\partial_j
\Gamma^{(\pm) k}_{il}+ \Gamma ^{(\pm) k}_{im} \Gamma ^{(\pm) m} _{jl}-
 \Gamma ^{(\pm) k}_{jm} \Gamma ^{(\pm) m} _{il}, \qquad R^{(\pm)}  _{ij}=
R^{{(\pm)}
k}{}_{ikj}\eqn\cur$$ The equation \conf\ can be obtained from varying the
action
$$S= \int d^D x \, e^{-2\Phi} \sqrt{|g|}\left( R- {1\over 3} H^2+ 4 (\nabla
\Phi)^2\right)
\eqn\acttar$$
The target space coordinates $x^i$ are the lowest components of the superfields
$\phi^i$ ($i=1,...,
D$).

The sigma model is invariant under (2,1) supersymmetry [\HW-\Van] if the target
space is even
dimensional ($D=2n$) with a complex
structure
$J^i{}_j$    which is covariantly constant
$$ \nabla ^{(+)}_k J^i{}_j=0\eqn\covj$$
with respect to the connection $\Gamma^{(+)}$ defined in \con, and with respect
to which the metric is hermitian, so that
$J_{ij}\equiv g_{ik}J^k{}_j$ is antisymmetric.

It is useful to introduce complex coordinates $z^\alpha, \bar z^{\bar \beta}$
in which the line element is
$ds^2=2 g_{\alpha \bar \beta}d z^\alpha d \bar z^{\bar \beta}$ and consider the
Dolbeault
 cohomology. An $N$-form is decomposed
into a set of $(p,q)$ forms with $p$    factors of  $dz$  and $q$  factors of
$d\bar z$ with $p+q=N$.
The exterior derivative decomposes as $d=\partial +\bar \partial$ and it is
useful to define $\hat d
=i (\partial - \bar \partial)$
and $\Delta = i\partial \bar \partial  = {1\over 2} \hat d d$. As $\Delta
^2=0$, $\Delta$ defines its own cohomology. Useful
lemmas are (i) if $\partial U=0$ and $\bar \partial U=0$ for some $(p,q)$ form
$U$, then locally there is a $(p-1,q-1)$
form $W$ such that $U= \Delta W$ (ii) if $\Delta U=0$ for some for some $(p,q)$
form $U$, then locally there is a $(p-1,q )$
form
$W$ and a $(p,q-1)$ form $X$ such that $U= \bar \partial X+ \partial W$.

 The   conditions for (2,1) supersymmetry imply that the (0,3) and (3,0) parts
of the 3-form $H$
 vanish, and $H$ is given in
terms of the fundamental 2-form
 $$J={1 \over 2}J_{ij} d \phi ^i _ \Lambda d \phi ^j
=ig_{\alpha \bar \beta} dz^\alpha_\Lambda d \bar z^ {\bar \beta}\eqn\jis$$
by
$$H=i(\partial - \bar \partial) J\eqn\hiss$$
Then the condition $dH=0$
implies
$$ i\partial \bar \partial J=0\eqn\jcon$$
so that locally there is a (1,0) form $k=k_\alpha d z^\alpha$ such that
$$J= i (\partial \bar k + \bar \partial k)\eqn\jisk$$
The metric and torsion potential are then given, in a suitable gauge,  by
$$
\eqalign{ g_{\alpha \bar \beta}&= \partial _ \alpha \bar k_ {\bar \beta} +
\bar \partial _
{\bar \beta}
 k_ \alpha
\cr
 b_{\alpha \bar \beta}&= \partial _ \alpha \bar k_ {\bar \beta} -  \bar
\partial _  {\bar \beta}
 k_ \alpha
\cr}
\eqn\kgeom$$
If $k_\alpha = \partial_\alpha K$ for some $K$, then the torsion vanishes and
the manifold is Kahler with Kahler potential $K$,
but if  $dk \ne 0$ then the space is a hermitian manifold of the type
introduced in [\HW].  The metric and torsion are invariant
under [\mohab]
$$ \delta k_\alpha = i \partial_ \alpha \chi + \theta_\alpha\eqn\ksym$$
where $\chi$ is real and $\theta_\alpha$ is holomorphic, $\partial_{\bar \beta}
\theta_ \alpha=0$.
It will be useful to define the vector
$$v^i=H_{jkl}J^{ij}J^{kl}
\eqn\vis$$
together with  the $U(1)$ part of the curvature
$$C^{(+)}_{ij}=J^l{}_kR^{{(+)} k} {}_{lij}\eqn\cis$$
and the $U(1)$ part of the connection \con\
$$\Gamma^{(+)} _i=J^k {}_j\Gamma^{(+)j}_{ik}=i( \Gamma^{(+)\alpha}_{i \alpha}-
\Gamma^{+\bar \alpha}_{i\bar \alpha})\eqn\uip$$
In a complex coordinate system, \cis\ can be written as
$C^{(+)} _{ij}= \partial _i \Gamma^{(+)} _j - \partial_j \Gamma^{(+)} _i$.

If the metric is Riemanian, then the holonomy of any metric connection
(including $\Gamma^{(\pm)}$)
is contained in $O(2n)$, while if it has signature $(2n_1,  2n_2)$ where
$n_1+n_2=n$, it will be in
$O(2n_1,2n_2)$. The holonomy ${\cal H}(\Gamma^{(+)})$ of the connection
$\Gamma^{(+)}$ is contained
in $U(n_1,n_2)$. It will be contained in $SU(n_1,n_2)$ if in addition
$$C^{(+)} _{ij}=0\eqn\ciso$$
where the $U(1)$ part of the curvature is given by \cis.
As $C_{ij}$ is a representative of the first Chern class, a necessary condition
for
this is the
vanishing of the first Chern class.

It was shown in [\Hsig] that geometries for which
$$\Gamma^{(+)} _i=0\eqn\fieldeq$$
in some suitable choice of coordinate system  will satisfy
the one-loop conditions \conf, provided the dilaton is chosen as
$$ \Phi= -   \log | det g_{\alpha \bar \beta}|
\eqn\fiis$$
which implies
$$\partial _i \fff={1 \over 2} v_i
\eqn\fisv$$
 Moreover, the one-loop dilaton
field equation is also satisfied for compact manifolds, or for non-compact ones
in which $\nabla
\Phi$ falls off sufficiently fast [\Hsiga-\Van]. This implies that ${\cal
H}(\Gamma^{(+)})
\subseteq SU(n_1,n_2)$ and these geometries generalise the Kahler Ricci-flat or
Calabi-Yau
geometries, and reduce to these in the special case in which
$H=0$.

These are not the most general solutions of \conf. In the special case in which
$H=0$ and the geometry is Kahler, the condition
\conf\ becomes
$$R_{ij}=2\nabla_i \nabla_j \Phi
\eqn\fdsf$$
which implies that either $\Phi$ is constant and the geometry is
Kahler-Ricci-flat, or that $J^{ij}\nabla_j \Phi$ is a Killing
vector, and the geometry is a generalised \lq linear dilaton' vacuum of a type
that has been analysed in
[\Kou]. If $H\ne 0$, then this generalises and the solutions are either of the
type described above,
or are ones in which
\fiis\ isn't satisfied and which have a Killing vector; this latter case will
be discussed in
[\prep] and here we will restrict ourselves to the case of $SU(n_1,n_2)$
holonomy with
\fieldeq,\fiis\ holding.

The equation \fieldeq\ can be viewed as a field equation for the potential
$k_\alpha$.
It can be obtained by varying the action
$$
S=\int d^{D} x \sqrt{ | det g_{\alpha \bar \beta}}|
\eqn\act$$
where $g_{\alpha \bar \beta}$ is given in terms of $k_\alpha$ by \kgeom.
This action was obtained independently by Martinec and Kutasov [\MKA,\Mucsb].
It can be rewritten as
$$
S=\int d^{D} x |   det g_{ij}|^{1/4}
\eqn\acta$$
which is non-covariant, as the field equation \fieldeq\ was obtained in a
particular coordinate system.
However, it is invariant under volume-preserving diffeomorphisms.

\chapter{(4,1) Supersymmetry, Real Structures, Hypersymplectic Structures,  and
Kleinian Geometry}

It was argued in [\HPf] that (4,1) sigma-models are finite to all orders in
perturbation theory.
For Euclidean signature, the model \acton\ will have (4,1) supersymmetry
if the complex dimension is even,
$n=2m$, and
${\cal H}(\Gamma^{(+)} ) \subseteq USp(m)$ (with $USp(1)=SU(2)$).
This implies that there are three complex structures $ I,J,K $   satisfying the
quaternion
algebra
\quat\
and
each satisfying \covj:
$$\nabla ^{(+)}_k I^i{}_j=\nabla ^{(+)}_k J^i{}_j=\nabla ^{(+)}_k
K^i{}_j=0\eqn\covq$$
The algebra \quat\ can   be written as
$$J^a J^b =- \delta ^{ab} {\II}+ \ee ^{abc} J^c
\eqn\fgsfdg$$
where $a=1,2,3$ and $J^a=\{I,J,K\}$.

In particular, as $USp(1)=SU(2)$, it
follows that in four dimensions,
$D=2n=4$, a geometry satisfying  \fieldeq\ will be finite to all orders and so
there are no
corrections to the action \act\ of higher order in the sigma-model coupling
constant $\alpha '$.
The curvature is anti-self-dual, satisfying
$$R^{(+)} _{ijkl}=-\2 \ee _{ij}{}^{mn}R^{(+)} _{mnkl}
\eqn\sdu$$

Consider now the case of non-Euclidean signature.
For 4 dimensions with the Kleinian
signature (2,2), the vanishing of the $U(1)$ part of the curvature
 implies that the curvature is again anti-self-dual, \sdu, and that the
holonomy is
$SU(1,1)=SL(2,\R)=Sp(1,\R)$. There are no longer three complex structures
but there are three covariantly constant tensors
$ J,S,T$   satisfying the pseudoquaternion algebra [\klein,\hit]
\slquat\
or, equivalently,
$$S^a S^b =- \eta ^{ab} {\II}+f^{ab } {}_c S^c
\eqn\sdfh$$
where $a=1,2,3$, $S^1=J,S^2=S,S^3=T$, $\eta ^{ab}= diag (+1,-1,-1)$ is the
$SL(2,\R)$ Killing metric
and
$f^{ab } {}_c$ are the $SL(2,\R)$ structure constants.
Each of the $S^a$ is covariantly constant with respect to the connection $\ggg
^{(+)} $
$$\nabla ^{(+)}_k J^i{}_j=\nabla ^{(+)}_k S^i{}_j=\nabla ^{(+)}_k
T^i{}_j=0\eqn\covh$$
and each satisfies
$$S^a_{ij}=-S^a_{ji}
\eqn\therm$$
The complex structure is $J=S^1$ which squares to $-\dd ^i{}_j$
while $S^2,S^3$ each square to $+\dd ^i{}_j$.
Each of the $S^a$ is requireds to be integrable, so that the
Nijenhuis-type tensor vanishes and there is a coordinate system in
which the real or complex structure is constant.
\foot{The   case   of almost complex structures or almost real structures which
are not
integrable will not be considered here, although they do lead to more general
models.}
 However they are not simultaneously integrable in general
i.e. for each $S^a$  there is a coordinate system in which $ (S^a)^i{}_j$ is
constant, but there
may not be one in which all three are simultaneously constant.

The $S,T$ are each real structures [\klein,\hit], sometimes called locally
product structures
[\GHR,\Ya]. If the
$S^a_{ij}$ ($a=2,3$) had been symmetric, the metric would have been a locally
product metric and the
space would have been  a locally product space of the type discussed in [\GHR].
The fact that they
are anti-symmetric gives a different structure, however. Choosing adapted real
coordinates $u^\aa,
v^{\tilde \aa} $ ($\alpha=1,2;{\tilde \aa}=1,2$) in which
$S $, say, takes the form
$$S^i {}_j= \pmatrix{
\delta^\alpha {}_\beta & 0 \cr
0 & -\delta ^{\tilde \alpha}{}_{\tilde \beta}
 \cr
}
\eqn\ttis$$
the condition \therm\ implies that the line element takes the form
$$ds^2=2g_{\aa {\tilde \aa}}(u,v)du^\aa dv^{\tilde \aa}
\eqn\metform$$
so that $\pa/\pa u^\aa$ and $\pa / \pa v^{\tilde \aa}$ are null vectors.
In general, this coordinate system will be incompatible with the complex
structure $J$.

Spaces of $SL(2,\R)$ holonomy have two  spinors that are covariantly constant
with respect
to $\ggg ^{(+)}$, $\eee^
A$ ($A=1,2$) and these can be used to construct three covariantly constant
2-forms
$S^{(AB)}=\bar \eee^A \gg  _{ij} \eee^B$, which can be identified with the
$S^a$; this gives the simplest
way of obtaining the above results. The sigma-models with these target spaces
do not have the usual (4,1)
supersymmetry. They have three currents
$$j^a=\2 S^a_{ij} \psi^i \psi ^j
\eqn\curs$$
generating an $SL(2,\R)$
Kac-Moody algebra and four supercurrents
$$G^0=\gij \pa X^i \psi^j , \qquad G^a=S^a_{ij} \pa X^i \psi^j
\eqn\scurs$$
The currents $T,G^0,G^1,j^1$ generate an $N=2$ superconformal algebra (where
$T$ is the
energy-momentum tensor). The full set of right-handed currents
$\{T,G^0,G^a,j^a\}$ generate a non-compact twisted form of the (small)
$N=4$ superconformal algebra. The global limit
is a (4,1) superalgebra in which the four right-handed supercharges
$Q^A=\{Q^0,Q^a\}$ (with $A=0,1,2,3$) satisfy
$$\{ Q^A Q^B \}=\eta ^{AB} P
\eqn\erte$$
where $\eta ^{AB}= diag (1,1,-1,-1)$ is the $O(2,2)$ Killing metric  and $P$ is
the right-moving momentum.

A similar structure obtains for spaces of signature $(2n,2n)$ with holonomy
$Sp(n,\R)$ -- the subgroup
of $U(n,n)$ preserving an anti-symmetric matrix, or equivalently the subgroup
of $O(2n,2n)$
preserving three matrices $S,T,J$ satisfying \slquat. Sigma-models with such
target spaces will have twisted
$(4,1)$ supersymmetry. Spaces with  one covariantly constant real structure $S$
satisfying the  conditions discussed above will have
twisted (2,1) supersymmetry, with global limit
given by \erte\ with $A=1,2$ and $\eta ^{AB}= diag (1, -1)$. If the metric is
to be invertible ($det g \ne 0$), this   requires
the metric to have signature $(m,m)$ and the holonomy is then ${\cal H} (\Gamma
^{(+)}) \subseteq GL(m,\R)$.

 Consider first the case in which there is no torsion, $H=0$. Then the
antisymmetric tensors
$J^a_{ij}$ or $S^a_{ij}$ are closed, as a result of \covq\ or \covh,
and so each closed 2-form defines a symplectic structure.
For Euclidean signature, the metric is Kahler with respect to each of the
complex structures $I,J,K$
 and the space is hyperkahler. The $\{ I,J,K \}$ constitute a   hyperkahler
structure. In complex coordinates adapted to any one of the complex structures,
the metric is
$$ds^2 = 2 g_{\aa {\bar \beta}} dz^\aa d \bar z^{\bar \beta}, \qquad
g_{\aa {\bar \beta}}= {\pa^2 \over \pa z^\aa \pa \bar z^{\bar \beta}  }K
\eqn\kah$$
for some locally defined Kahler potential $K$.

For signature $(2n,2n)$, the $\{ J,S,T \}$ constitute a hypersymplectic
structure [\hit].
The metric is Kahler with respect to the complex structure $J$, while in
coordinates adapted to
either of the  real structures, $S$ say,
the metric takes the form
$$ds^2=2g_{\aa \bbt}(u,v)du^\aa dv^\bbt,
 \qquad
g_{\aa \bbt}= {\pa^2 \over \pa u^\aa \pa \bar v^\bbt }K
\eqn\metformh$$
for some locally defined potential $K$. In these coordinates, the symplectic
structure is
$S=g_{am}du^a_\lll dv^m$.

If $H \ne 0$, then the 2-forms $J^a $ or $S^a $ are not closed, but $I,J,K$ are
$\ddd $-closed.
In the Euclidean case,
one can choose complex coordinates adapted to any one of the three complex
structures, and the
 formulae \jis-\kgeom\ then hold for each choice of complex structure. For the
Kleinian signature
$(2n,2n)$, the complex structure
$J$ again leads to conditions \jis-\kgeom. For the real structure $S$ (or $T$)
it is
useful to introduce the adapted coordinates
$u^\aa,v^\bbt$,
and consider the analogue of Dolbeault cohomology. An $N$-form is decomposed
into a set of $(p,q)$ forms with $p$    factors of  $du$  and $q$  factors of
$dv$ with $p+q=N$.
The exterior derivative decomposes as $d=\partial_u +  \partial_v$
where $\pa _u:H^{(p,q)} \to H^{(p+1,q)}$ and $\pa _v:H^{(p,q)} \to
H^{(p,q+1)}$.
It is useful to define $\hat
d =  (\partial _u -  \partial _v)$
and $\Delta = \partial _u \partial  _v= {1\over 2} \hat d d$. Again $\Delta
^2=0$, so that $\Delta$ defines its
own cohomology.
 Then $H$ is given in
terms of the fundamental 2-form
 $$S={1 \over 2}S_{ij} d \phi ^i _ \Lambda d \phi ^j
= g_{\aa \bbt}du^\aa _ \Lambda dv^\bbt\eqn\sis$$
by
$$H= (\partial _u-   \partial _v) S\eqn\hisss$$
The   condition $dH=0$ then
implies
$$  \partial _u \partial _v S=0\eqn\jcon$$
so that locally there is a (1,0) form $k=k _\aa d u^\aa$ and a (0,1) form
$\tilde k=\tilde k_ \bbt dv^\bbt$ such that
$$S= \partial _u \tilde k +  \partial _v k\eqn\abcs$$
The metric and torsion potential are then given, in a suitable gauge,  by
$$
\eqalign{ g_{\aa \bbt}&= \partial _ \aa   \tilde k_ \bbt +   \partial _  {\bbt}
 k_ \aa
\cr
 b_{\aa \bbt}&= \partial _ \aa  \tilde k _ \bbt -   \partial _  {\bbt}
 k_ \aa
\cr}
\eqn\kgeom$$
so that
$$H= \pa _u \pa _v (k+\tilde k)
\eqn\abc $$
If $k_\aa = \partial_\alpha \kk$ and $\tilde k_\bbt= \pa _\bbt \tilde \kk$ for
some
locally defined potentials $\kk,\tilde \kk$, then
the torsion vanishes and
$$S= \pa _u \pa _v (\tilde \kk- \kk)
\eqn\abc $$
so that \metformh\ is satisfied with potential $K=\tilde \kk- \kk$.

The power-counting arguments of Howe and Papadopoulos [\HPf]
can be generalised to   apply
to models with this twisted $(4,1)$ supersymmetry, so that
such models should again be finite.
 This is supported by the results of   Martinec and Kutasov [\MKA], who showed
that the action \act\
generates the correct S-matrix for part of the (2,1) string,
 confirming that this action receives no
corrections in $D=4$.

For target spaces of signature $(m,m)$ with holonomy $GL(m,\R)$ with one
covariantly constant integrable  real structure $S$
satisfying \covh,\therm,  the  geometry is given in terms of a scalar potential
by \metformh\ if
$H=0$ or  by a vector potential \kgeom\ if $H \ne 0$. The metric and torsion
are preserved by the
gauge transformations
$$\dd k_\aa = \pa _\aa \chi + \th _\aa, \qquad
\dd \tilde k_\aat =- \pa _\aat \chi +\tilde  \th _\aat
\eqn\abc $$
where
$\pa _ \aat \th _\aa= \pa _\aa \tilde  \th _\aat
=0$.
In analogy with \vis,\cis,\uip,
it will be useful to define the vector
$$\tilde v_i=H_{ijk}S^{jk}
\eqn\vtis$$
together with  the $GL(1,\R)$ part of the curvature
$$\tilde C^{(+)}_{ij}=S^l{}_kR^{{(+)} k} {}_{lij}\eqn\ctis$$
and the $GL(1)$ part of the connection \con\
$$\ti \Gamma^{(+)} _i=S^k {}_j\Gamma^{(+)j}_{ik}=( \Gamma^{(+)\alpha}_{i
\alpha}-
\Gamma^{+\ti \alpha}_{i\ti \alpha})\eqn\utip$$

If $H=0$, then the curvature 2-form is a (1,1) form and the only non-vanishing
components of the
curvature are $R_{\aa \bbt \gg \ti \dd}$. It follows that the Ricci tensor
$R_{\aa\bbt}$ is
proportional to is proportional to $\ti C_{\aa\bbt}$
and is given by
$$R_{\aa\bbt}= \pa _\aa \pa _\bbt \log |det g_{  \gg \ti \dd}|
\ek
with $R_{\aa\bb}=0$.
Thus the Einstein equation $R_{ij}=0$ is equivalent to demanding $SL(m,\R)$
holonomy and
gives, with a suitable choice of coordinates,
$$|det g_{  \gg \ti \dd}|=1
\ek
which gives    a Monge-Ampere equation for $K$ on using \metformh.

If $H \ne 0$,
the condition  \fieldeq\ of the complex case is replaced by
$$\ti \Gamma^{(+)} _i=0
\eqn\fieldeqt$$
and this again  implies that
the one-loop field equation \conf\ is satisfied, provided the dilaton is chosen
as \fiis.
Furthermore, the condition \fieldeqt\ implies
$\ti C^{(+)}_{ij}=0$ and so the holonomy is in $SL(m,\R)$.
The field  equation \fieldeqt\ can again be obtained from the action \acta, but
where now the metric
is given by \kgeom\ in terms of the potentials $k, \ti k$ corresponding to the
real structure $S$,
and it is these that are varied to give the field equation \fieldeqt.

It is remarkable how much of the geometry based on a complex structure $J$
carries over to the case of a real structure $S$.
Instead of using complex numbers, it is sometimes useful to use {\it double
numbers} in this context, which are based on introducing a number
$e$ satisfying $e^2=1$ instead of the usual imaginary unit $i$ satisfying
$i^2=-1$ [\klein].

 \chapter{The (2,0)-Supersymmetric Sigma-Model and the Bott-Chern Form}

Consider now the (1,0) sigma-model. It consists of (1,0) scalar superfields
$\fii$ taking values in
the target space $M$ and coupling to
$\gij$ and $\bij$, plus fermionic fields $\psi ^M$ which are sections of
$S_+\times V$ where $S_+$
is the world-sheet chiral spinor bundle and $V$ is a vector bundle over $M$
with structure group
$G$; they couple to the connection $A_i $ on $V$ [\HW]. The (1,0) superspace
action is [\HW]
$$S=\int d^2x d  \th \, \left( \gij +\bij \right) \pa _- \ffi ^i D \phi ^j +
\psi ^M (D \psi ^M
 + A_i^{M }{}_N D\ffi \, \psi ^N)
\eqn\sigt $$
where $D$ is the superspace supercovariant derivative
The conditions for   conformal invariance are   derived from the action
$$\eqalign{
S= \int d^D x e^{-2\Phi} \sqrt{|g|}
&\left(  R- {1\over 3} H^2+ 4 (\nabla \Phi)^2
\right.
\cr
&
\left.
-{\alpha'  \over 2} [\tr(F_{ij}F^{ij})-R^{(-)}_{abij}R^{(-)baij}]+
O(\alpha '{}^2)\right)
\cr}
\eqn\acttar$$
where $H$ is now given by
$$
H = {1 \over 2}db+ \alpha'  [\Omega (A)-\Omega (\omega^-)]
\eqn\hcs$$
and $\Omega$ is the Chern-Simons 3-form
$$\tr (F^2)=d \Omega (A), \qquad \Omega (A)=\tr(AdA+ {\textstyle{2\over 3}}
A^3)\eqn\csis$$
The curvatures $R^{(\pm)}$ and connections $\Gamma^{(\pm)}$ are given by
\con,\cur\ with the torsion \hcs. A
vielbein $e_i^a$  has been introduced, with   the corresponding spin
connections
$\omega_i^{(\pm) ab}$, curvatures $R^{(\pm)
a}{}_{bij}$ and curvature 2-forms $R^{(\pm)
a}{}_{b }$.
The gravitational Chern-Simons term is given by
$$\tr(R^{(\pm) 2})= R^{ (\pm) a}{}_b R^{ (\pm) b}{}_a
=d \Omega (\omega^{(\pm)}), \qquad \Omega (\omega^{(\pm)})=\tr(\omega^{(\pm)}
d\omega^{(\pm)}+{\textstyle{2\over 3}}
\omega^{(\pm) 3})\eqn\grcs$$
The new torsion satisfies
$$dH=\alpha'  [\tr(F^2)-\tr (R^{(-) 2})]\eqn\kida$$
and the condition
$$\int _\ss \tr (F^2)= \int_\ss \tr (R^{(-) 2})
\eqn\glob$$
is required over any 4-cycle $\ss$ for $H$ to be well-defined.
A key identity is
$$R^{(+)}_{ijkl}-R^{(-)}_{klij}= -2H_{[ijk,l]}
\eqn\kid$$
which can be rewritten using \kida.
As $H$ appears in the gravitational Chern-Simons term  on the right hand side
of \hcs,
the equations \hcs,\grcs\ only implicitly define $H$, but $H$ can be
constructed perturbatively
in $\alpha '$.

The model has (2,0) supersymmetry classically if (i) $(M,\gij,\bij)$ is a (2,1)
geometry, i.e. a
hermitian space with torsion whose complex structure is covariantly constant
\covj\ with respect to
the connection $\ggg^{(+)} $ defined by \con,\hcs,
 and (ii) $V$ is a holomorphic vector bundle, i.e. the field strength
$F=dA+A^2$ is a (1,1) form
[\HW]. This implies that the (1,0) part of the connection ${\cal A}=A_\alpha
dz^\aa $ satisfies
$${\cal A}=V^{-1}\pa V\eqn\ais$$ for some complex $G$-valued function $V$, i.e.
$V$ takes values in
the complexification of $G$.  (A group element in a neighbourhood of the
identity is of the form
$g=\exp {\aa _m t^m}$ where $\aa _m$ are real coordinates and $ t^m$ are
elements of the Lie algebra
${\cal g}$. The prepotential is of the form $V=\exp {v _m t^m}$ where $v_m$ are
complex, and $\ba
V=\exp {\ba v _m t^m}$.)  Under a  gauge transformation with parameter $g (x)
\in G$
$$ A \to g ^{-1} d g + g^{-1} A g
\eqn\gagvar$$
As $A={\cal A}+{\cal A}^*$,
the connection will
be pure gauge if
$V$ is real.
The prepotential $V$ transforms as
$$V \to \bar  \lambda V g
\eqn\vvar$$
under a gauge transformation   and under a pre-gauge transformation
with holomorphic $G$-valued  parameter
$\ll (z)
\in G$; the pre-gauge transformations leave ${\cal A}$ invariant. It is also
useful to define
$$U=V \bar V^{-1}
\eqn\abc $$
which is invariant under the gauge transformations since $g$ is real, but
transforms under the
pre-gauge transformations as
$$ U \to  \bar \lambda U\lambda ^{-1}
\eqn\uvar$$

The gauge transformations \gagvar,\vvar\ have  parameter $g$ taking values in
$G$.
Consider the
(1,0) form
 $$
a = U^{-1} \partial U
\eqn\abc $$
which can be rewritten as
$$  a = \bar V A \bar V ^{-1} + \bar V d \bar V ^{-1}
\eqn\abc $$
Thus $a$ is related to $A$ by a  complex gauge transformation
  \gagvar,\vvar\ with parameter $g= \bar V^{-1}$ taking values in the {\it
complexification } of
$G$.
Thus the complexified vector bundle $V_c$ is a holomorphic bundle with
holomorphic connection $a$
(see e.g. [\HP]).
Similarly, the complex gauge transformation
$$A \to \bar a \equiv V A  V ^{-1} +   V d  V ^{-1}=   U \bar \partial U ^{-1}
\eqn\abc $$
defines an anti-holomorphic connection $\bar a$ which is a (0,1) form.
Under the pre-gauge transformations
\uvar,
$$ a \to
\lambda^{-1}a \lambda + \lambda ^{-1} \partial \lambda
\eqn\abc $$
and the field strength is
$$f=da-a^2 = \bar \partial a
\eqn\abc $$
since the (0,1) part of $a$ vanishes.
This is related to $F$ by
$$ F= \bar V f \bar V ^{-1}= V \bar f V ^{-1}
\eqn\abc $$
so that
$$ \tr  (F^n)=\tr  (f^n)=\tr  (\bar f^n)
 \eqn\abc $$

 As the (2,2) form $\tr (F^2)$ satisfies $\partial \tr (F^2)=\bar \partial \tr
(F^2) =0$, then by lemma
(i) there is a (1,1)
form $\Upsilon(V,\bar V)$ such that
$$\tr (F^2)= i \partial \bar \partial \Upsilon
\eqn\bcf$$
$\Upsilon(V,\bar V)$ is the Bott-Chern 2-form [\Bch], constructed  in
[\Bcha,\Don,\HP].
The Chern-Simons form $\Omega (A)$ given by \csis\ then satisfies
$$\Omega(A)= \hat d \Upsilon + d \chi
\eqn\csbc$$
for some 2-form $\chi(V,\bar V)$.
Note that the Bott-Chern form can be written entirely in terms of $U$,
$\Upsilon(V,\bar V)=\Upsilon(U)$ but $\cc$ cannot.

An instructive example is that in which $G$ is abelian.
Then $F^m=dA^m$ and there are real scalars $\phi ^m, \theta^m$ ($m=1,\dots,
rank(G)$) such that
$$\eqalign{
A^m &={\cal A}^m +\bar {\cal A}^m= d \th ^m + \hat d \phi ^m, \qquad {\cal
A}^m=\partial
(\theta^m+i\phi ^m),
\cr  a^m
&=2i \pa \phi^m \qquad  \bar a^m=-2i \bar \pa \phi^m
\cr}
\eqn
\abd$$
and
$$V=\exp (\theta +i\phi ), \qquad U=\exp ( 2i\phi)
\eqn\viss$$
Then
$$
F^m=dA^m = \pa a = \bar \pa \bar a =-2i\partial \bar
\partial
\phi^m
\eqn\abc $$
and
$$\tr (F^2)=-4 \partial \bar \partial \phi^m\partial \bar \partial \phi^m$$
and the Bott-Chern form can be chosen to be
$$\Upsilon=-4i \partial   \phi^m  \bar \partial \phi^m$$
The Chern-Simons form $AdA$ then satisfies \csbc\ with
$$\chi =-2i \th ^m F^m
\eqn\gdtgd$$
Under a gauge transformation \gagvar,\vvar\
with $g =e^\aa $ and $\ll = e^{2l}$ (with $\bar \pa l^m=0$)
$$\dd A ^m_i=\pa _i \aa ^m, \qquad \dd \ff ^m =i(l^m-\bar l^m) ,\qquad \dd \th
^m = \aa ^m +(l^m
+\bar l^m)
\eqn\ggtr$$

In the non-abelian case,
introducing coordinates $\phi ^m$ on the group manifold $M$, one has
$$\uuu = (G_{mn}+B_{mn}) \pa \phi^m \bar \pa \phi ^n
\ek
defining a metric $G_{mn}(\ffi)$ and anti-symmetric tensor $B_{mn}(\ffi)$.
This can be constructed explicitly as follows [\Don].
Let $A(t,x^\mm)$ be a 1-parameter family of connections labelled by
 $0\le t\le 1$, constructed from pre-potentials $V(t,x^\mm)$ with corresponding
$t$-dependent $U,a,f,F$ defined as above.
Then
$${\pa \over \pa t}   f=\ba \pa
\dot a= \ba \pa  \pa _a (U^{-1} \dot U )
\ek
where
$$
\pa _a (U^{-1} \dot U )\equiv \pa   (U^{-1} \dot U ) +[a,U^{-1} \dot U]
\ek
so that
$$ \eqalign{{\pa \over \pa t} \tr  F^n&={\pa \over \pa t} \tr  f^n=
n \, \tr  \left(\ba \pa _a (U^{-1} \dot U ) f^{n-1}
\right)
\cr &
=n \ba \pa \pa _a \tr  \left( (U^{-1} \dot U ) f^{n-1}
\right)
\cr &
=n \ba \pa \pa  \tr  \left( (U^{-1} \dot U ) f^{n-1}\right)
\cr}
\ek
Thus if $F(1,x^\mm)=F(x^\mm)$ and $F(0,x^\mm)=\hat F(x^\mm)$,
$$\tr (F^n)=\tr (\hat F^n)+ i\ba \pa \pa \uuu_n
\ek
where
$$\uuu_n= -i{1 \over n}\int dt \, \tr  (U^{-1}\dot U f^{n-1})
=-i{1 \over n}\int dt \, \tr  (U^{-1}\dot U [\ba \pa (U^{-1}\pa U)]^{n-1})
\eqn\bcis$$
The case $n=2$ defines the form needed here,  $\uuu(U)=\uuu_2$, which will
exist locally.
Note that it is only defined by \bcf\
up to the addition of  a $\ddd$-closed term, $\uuu \to \uuu + \pa X+\ba \pa Y$.

In four dimensions, the Donaldson action
$$\int J _\lll \uuu
\eqn\don $$
gives an action on any hermitian space with complex structure 2-form $J$ whose
variation with respect to $U$ or $\ffi$ implies that $F$ is
self-dual. In $2n+2$ dimensions, the action
$$\int J^n {}_\lll \uuu
\eqn\abc $$
implies that the (2,0) part of $F$ vanishes, and $F$ satisfies the
Uhlenbeck-Yau equation
$$ J^{ij}F_{ij}=0\eqn\uyau$$
The two-dimensional case $S= \int \uuu$ gives a Wess-Zumino-Witten model for
the complexification of $G$.

For geometries in which $dH=0$ (e.g. for the (2,1) sigma-model or for the (2,0)
model
in the classical   limit $\alpha'\to 0$), the fact that $\omega ^{(+)}$ has
$U(n_1,n_2)$ holonomy
together with \kid\ implies that  $R ^{(-)}$ is a
(1,1) form, so that the tangent bundle $T(M)$ with connection
 $\omega ^{(-)}$ is holomorphic. Then there are complex $U(n_1,n_2)$-valued
scalars $W$ such that
$$ \omega ^{(-)}= W^{-1} \partial W +( W^{-1} \partial W )^*$$
and there is a Bott-Chern form $\Upsilon (Y)$ and a 2-form $\chi(W,\bar W)$
such that
$$\Omega(\omega ^{(-)})= \hat d \Upsilon + d \chi
\eqn\omcs$$
where
$$Y=W \ba W^{-1}
\ek
In the quantum case, the Chern-Simons corrections to $H$ and hence to $\omega
^{(-)}$ give $\alpha '$ corrections to these
equations, but again there are forms $\Upsilon (Y)$ and   $\chi(W,\bar W)$
satisfying \omcs\
which can be constructed
order by order in $\alpha '$.

There will be (2,0) supersymmetry in the quantum theory if
the complex structure is covariantly constant with respect to the
connection
given by
\con,\hcs, whose torsion
now includes the
Chern-Simons terms \hcs\ [\HW]; thus
  this   connection has $U(n_1,n_2)$ holonomy.
This  again implies that $H$ is given by \hiss, but now \kida\ implies
[\Hsiga,\Hsigb]
$$i \partial\bar \partial J= \alpha'  [\tr (F^2)-\tr (R^{(-) 2})]
\eqn\jeq$$
This implies the local existence of a (1,0) form $k$ such that
$$J=\alpha'  \hat \Upsilon + i (\partial \bar k + \bar \partial k)\eqn\abc$$
where
$$\hat \Upsilon=\Upsilon(U)-\Upsilon(Y)
\ek
which will be well-defined if \glob\ holds.
Then the metric is given by
$$
 g_{\alpha \bar \beta} =\partial _ \alpha \bar k_ {\bar \beta} +  \bar \partial
_  {\bar
\beta}
 k_ \alpha + i \alpha'   \hat\Upsilon _{\alpha \bar \beta}
\eqn\metiss$$
while the torsion potential can be chosen to be
$$
  b_{\alpha \bar \beta} =\partial _ \alpha \bar k_ {\bar \beta} -  \bar
\partial _  {\bar \beta}
 k_ \alpha
+ \alpha' \hat \chi _{\alpha \bar \beta}
\eqn\kgeom$$
where
$$\hat \chi _{\alpha \bar \beta}=
 \chi _{\alpha \bar \beta} (V,\bar V)-  \chi _{\alpha \bar \beta} (W,\bar W)
\ek
This is in agreement with the results of Howe and Papadopoulos [\HP], in which
it was shown that all
anomalies in the (2,0) sigma-model can be cancelled by adding finite local
counterterms to the
$\gij, \bij$, so that
$$
\gij \to \gij +\alpha'  \hat \Upsilon_{ij}, \qquad \bij \to \bij +\alpha'  \hat
\chi_{ij}
\eqn\shift$$
together with attributing to $\bij$ the standard anomalous transformations
$$\delta \bij=
\alpha ' \tr (Ad \alpha- \omega^{(-)} d \Lambda)
\ek
under Lorentz and gauge symmetries with parameters $\Lambda, \alpha$
respectively
[\HW,\HP].
Note that whereas shifting $\gij$ by a counterterm proportional to $\tr  A_i
A_j$, which was used in the
arguments of [\MKA], is sufficient to remove the sigma-model anomalies in the
(1,0) model, this is not
consistent with (2,0) supersymmetry and it is necessary to use the counterterms
\shift, as shown in
[\HP].

The Yang-Mills field equation is, to lowest order in $\ap$,
$$D^{(+) i}F_{ij}- 2\nabla ^i\Phi F_{ij}=0\eqn\ymeq$$
where $D^{(+)}$ is the gauge and gravitational covariant derivative involving
the connections  $\ggg^{(+)}$ and $A$.
This can be integrated to give the Uhlenbeck-Yau equation
 \uyau, which can be rewritten as
$$g^{\aa{\bar \beta}}F_{\aa{\bar \beta}}=0\eqn\abc$$
Indeed, differentiating \uyau\ and using \covj,\vis,\fisv\ gives \ymeq.
The Uhlenbeck-Yau equation \uyau\      will receive higher order corrections in
$\ap$ in general.
Note that the complex structure  $J^{ij}$ in \uyau\ is the modified one
containing the Bott-Chern
form
$\hat \uuu$.

The conditions given above are sufficient for the sigma-model to be
conformally invariant to lowest order in $\ap$. These are not the
most general solutions, but they are precisely the ones
that will admit   Killing spinors and so be invariant under   spacetime
supersymmetries when considered as superstring
backgrounds [\Hsiga,\Hsigb].
 The more general backgrounds, which necessarily have an isometry, will be
discussed in
[\prep]. These field equations are obtained by varying  the action
\acta\ with respect to $k_\aa$ and $V$, where $\gab$ is given by \metiss. Then
\acta\ is the
effective action generating the conformal invariance conditions for (2,0) sigma
models to lowest
order in $\aa '$, and so is the leading part of the effective action for (2,0)
strings.

Consider now the conditions for the (1,0) action \sigt\
to have a twisted (2,0) supersymmetry. As  in the last section, this requires
the existence of a
real structure $S$ on $M$
satisfying \covh,\therm.
Invariance of the terms in \sigt\ involving the fermionic superfields $\psi $
requires that the
Yang-Mills field strength satisfies
$$S_{[i}{}^kF_{j]k}=0\eqn\dsfs$$
 so that the field strength  $F$ is a (1,1) form ($F_{\aa \bb}=0$, $F_{\aat
\bbt }=0$)
  and this implies that
$$A= {\cal A}+\tilde {\cal A}
\eqn\ait$$ where
$${\cal A}= V^{-1}\pa _u V , \qquad \tilde {\cal A} =\ti V^{-1}\pa _v \ti V
\eqn\aish $$
for two independent real potentials $V, \ti V$, each taking values in $G$ (not
its
complexification). The potential $A$ will be pure gauge if $V=\ti V$. The
Uhlenbeck-Yau equation
is replaced by
$$ S^{ij}F_{ij}=0\eqn\usyau$$
which is equivalent to
$$g^{\aa \bbt}F_{\aa \bbt}=0
\ek
The results described above for the usual  (2,0) model generalise
straightforwardly to this
twisted case. In particular, there is a Bott-Chern-type form $\ti \uuu$ and a
form $\ti \chi$
such that
$$\tr (F^2)=  \ap \partial _u   \partial _v \ti \Upsilon
\eqn\bcfh$$
 and the Chern-Simons form $\Omega (A)$   \hcs\ is given by
$$\Omega(A)= (\pa _ u - \pa _v)\ti  \Upsilon (U)+ d \ti \chi (V,\ti V)
\eqn\csbch$$
where $U=V\ti V^{-1}$.
Similarly, the spin-connection has prepotentials $W,\ti W$ and the
gravitational Chern-Simons term gives a form $ \ti \Upsilon (Y)$ with $Y=W\ti
W^{-1}$.
and the quantum metric is
$$
 g_{\alpha \ti \beta} = \ap \hat\Upsilon _{\alpha \ti \beta}+\partial _ \alpha
\ti k_ {\ti\beta} +  \ti \partial _  {\ti
\beta}
 k_ \alpha
\eqn\metissh$$
where
$$\hat\Upsilon=\ti  \Upsilon (U)-\ti  \Upsilon (Y)
\ek
The field equations are again obtained by varying \acta, where the metric is
given by
\metissh.

 \chapter{The (2,1) String}

For the (2,0) sigma-model to have (2,1) supersymmetry, it is necessary that the
fermions $\psi ^M$ split into a set $\psi^i=e^i_a \psi ^a$ which can combine
with the (2,0) superfields $\ff^i$  to form (2,1)
supermultiplets, and a set $\psi^{M'}$ on which the extra supersymmetry is
non-linearly realised.  Thus the vector bundle $V$
should be   of the form  $TM\times V'$ where TM is the tangent bundle and V' is
 some other bundle with structure group $G'$ [\MK].
The structure group $G$ of $V$ is then in $G'\times O(n)$.
The fermions $\psi^M$ then split into $(\psi^a, \psi^{M'})$, with $M'=1, \dots
dim (V')$. The $\psi^a$   are sections of
$TM\times S_+$ and are the superpartners of the (2,0) scalar multiplets. The
supercurrent
generating the extra supersymmetry on the fermions $\psi^{m'}$
is of the form
$$G=\6
f_{M'N'P'}\psi^{M'}\psi^{N'}\psi^{P'}\eqn\abc$$
where $f_{M'N'P'}$ are the structure constants of some Lie group, so that this
supersymmetry is
realised non-linearly on the fermions.

The connection $A$ decomposes into a connection on $TM$ and a connection $A'$
on $V'$.
The   connection on TM given by restricting the $V$ connection  $A$ to $TM$
must be
gauge-equivalent to $\ww ^{(-)}$ [\Hsig-\Van], so that in a suitable gauge
$A=A' +\ww ^{(-)}$ and there is a
prepotential $V'$ for $A'$. Substituting this in the conditions obtained above
for
  (2,0) supersymmetry, we obtain
$$
H = {1 \over 2}db+\ap  \Omega (A')
\eqn\hcsp$$
As there are no gravitational Chern-Simons terms, $H$ does not appear on the
right hand side, so
that \hcsp\ gives $H$ explicitly.
The global condition \glob\ now becomes $\int_\ss \tr (F'F')=0$ for all
4-cycles $\ss$.
Then
$$i \partial\bar \partial J= \ap \tr (F')^2 \eqn\jeqp$$
and there is  (1,0) form $k$ such that
$$J= \ap  \Upsilon (U')+ i (\partial \bar k + \bar \partial k)\eqn\abc$$
and the metric and torsion potential  are given by
$$\eqalign{
 g_{\alpha \bar \beta} &= i  \ap\Upsilon _{\alpha \bar \beta}(U')+\partial _
\alpha \bar k_
{\bar \beta} +
\bar \partial _  {\bar
\beta}
 k_ \alpha
\cr
 b_{\alpha \bar \beta} &= i  \ap \chi _{\alpha \bar \beta}(V',\bar V')+\partial
_ \alpha \bar k_ {\bar
\beta} -
\bar \partial _  {\bar
\beta}
 k_ \alpha
\cr}
\eqn\erter$$
The Yang-Mills   equation becomes
$$ J^{ij}F'_{ij}=0\eqn\uyau$$
These equations can be obtained by varying the action \acta.

It will be useful to write the metric in terms of a fixed background metric
$\hat g_{\aa\bbb}$ (e.g. a flat metric)
which is given in terms of a potential $\hat k$ by
$\hat g_{\aa \bbb}= \pa _\aa \hat  k_ \bbb +\pa _\bbb \hat  k_\aa$, and a
fluctuation given in terms of  a vector field $B_i$ defined by
$$ B_\aa = -i (k_\aa-\hat k_\aa), \qquad
B_\ab = i(\bar k_\ab-\hat   k_\ab^*)
\eqn\abc $$
with field strength ${\cal F} =dB$. Then
$$   g_{\aa\bbb}=
\hat g_{\aa\bbb} +i{\cal F}_{\aa\bbb} +\ap \uuu _{\aa\bbb}
\eqn\abc $$
The gauge
  symmetry \ksym\ has become  the usual gauge transformation of an abelian
gauge field
$$\dd B_i = \pa _ i \chi
\eqn\tihrue$$
and the action \act\
becomes
$$
S=\int d^{D} x
\sqrt{ | det ( \hat g_{\aa\bbb} +i{\cal F}_{\aa\bbb} +\ap \uuu _{\aa\bbb} )|}
\eqn\actbf$$
which is similar to a
  Born-Infeld action. Note that
the (2,0) part of ${\cal F}$ is non-zero.
Note that
$$*det g_{\aa\bbb} \prop  J_\lll J = J^0 _\lll J ^0 + 2\ap J_\lll \uuu +(\ap)^2
\uuu _\lll \uuu
\eqn\abc $$
where $J^0 = J- \alpha ' \uuu = \pa \ba k + \ba \pa k$ is  the classical
complex structure, so that the
expansion of the action
\act\ includes a Donaldson term \don, plus other terms such as $\uuu^2$. These
are needed to ensure
that the connection is holomorphic and satisfies the Uhlenbeck-Yau equation
with respect to the
quantum complex structure $J$ which has gauge-field dependence, instead of with
respect to the
classical complex structure $J^0$.

Instead of   the usual (2,1) string or sigma-model, one can construct a string
or sigma-model based
on the twisted form of the (2,1) algebra. Much of the analysis is similar to
that for the usual
(2,1) string, but different  factors of $i$ and $-1$. There is a real structure
$S$ and the  classical metric is
\kgeom\ and the gauge potential $A'$ is given by \ait,\aish\
in terms of pre-potentials
$V',\ti V'$ for $A'$.
In the quantum case, the 2-form $S$ is
$$S= \alpha ' \ti \Upsilon (V',\ti V')+  \partial _u \ti  k +  \partial _v k
\eqn\abc$$
and the metric and torsion potential  are given by
$$\eqalign{
 g_{\alpha \ti \beta} &= \alpha ' \ti \Upsilon _{\alpha \ti \beta}(V',\ti
V')+\partial _ \alpha \ti k_
{\ti \beta} +
\ti \partial _  {\ti
\beta}
 k_ \alpha
\cr
 b_{\alpha \ti \beta} &= \alpha '\chi _{\alpha \ti \beta}(V',\ti V')+\partial _
\alpha \ti k_ {\ti
\beta} -
\ti \partial _  {\ti
\beta}
 k_ \alpha
\cr}
\eqn\erterh$$
The Yang-Mills   equation becomes
$$ J^{ij}F'_{ij}=0\eqn\uyau$$
and these equations can also be obtained by varying the action \acta.

For target spaces of signature (2,2)   with $SL(2,\R)=SU(1,1)$ holomnomy, the
sigma-model has twisted (4,1) supersymmetry which contains both the usual (2,1)
algebra and the
twisted one, and both approaches give the same result, but in terms of
different variables. The action in either approach is \acta, but can be viewed
as depending on the variables $k_\aa,k_\aab,U'$ through \erter\ or on
$k_\aa,k_\aat,\ti U'$ through \erterh.

For the action based on the twisted (2,1) formalism, it is useful to define
$B$ now by
$$ B_\aa =  ( k_\aa-\hat k_\aa), \qquad
B_\aat = -(\ti k_\aat -\hat  k_\aat)
\eqn\abc $$
in terms of the potential $k,\ti k$  of the twisted (2,1) sigma-model.
The gauge symmetry is again \tihrue\
and the
metric is now given by
$$   g_{\aa\bbb}=
\hat g_{\aa\bbb} -{\cal F}_{\aa\bbb} +\alpha '\uuu _{\aa\bbb}
\eqn\abc $$
This formalism based on the real structure
may be better suited to performing a null reduction with respect to a a null
Killing vector, as coordinates could be chosen so that the
Killing vector represents translation in one of the null coordinate directions
e.g. $\pa/\pa u^1$.


\chapter {The Schild Action}

The action \acta\ is of the form
$$S_{NG}=\int
d^{D} x \,  g^{\nu} \eqn\sac$$
where
 $g=|\det g_{ij}|$ and $\nu =1/4$, whereas the covariant Nambu-Goto action is
given by \sac\ with $\nu = 1/2$.
Instead of introducing an intrinsic  metric on the world-volume to obtain a
Polyakov-type action, one can introduce a scalar
world-volume gauge field $V$  to obtain a Schild-type action
$$S_{S}={1\over
2\nu}  \int d^{D}x\
\ V^{1-2\nu}g^{\nu}-(1-2\nu)\beta V{}\eqn\abc$$ where $\nu,\beta$ are
constants.
This
action is invariant under world-sheet diffeomorphisms:
$$\delta\phi =\xi^{a}\partial_{a}\phi ,\ \ \ \ \ \ \ \
\delta V=\partial_{a}(V\xi^{a})\eqn\dif$$ so that $V$ is a scalar density.

The $V$ field equation is
$$V= \beta ^{1\over2\nu}\sqrt{g}\eqn\abc$$
Substituting for $V$  in
the Schild action gives the Nambu-Goto action $$S_{S}\rightarrow\
\beta\mathop{}\nolimits^{2\nu-1\over2\nu}\int d^{D}x\ \sqrt{g}\eqn\abc$$
 so that these
two models are classically equivalent.

The diffeomorphism symmetry \dif\ can be partially fixed by imposing the gauge
condition $V=a$ to the Schild action,
where $a$ is some constant. The   gauge-fixed  lagrangian
 is the   Eguchi-type Lagrangian
$$L_E=g^{\nu}+C\eqn\abc$$
where $C$ is a constant.
The field equation $$\partial L_E/\partial\phi=0\
\Rightarrow \ g=\hbox{const}\eqn\abc$$
The Eguchi action is invariant under the volume preserving
diffeomorphisms
 $$\delta\phi^{\mu}=\xi^{a}\partial_{a}\phi^{\mu},\
\ \partial_{a}\xi
^{a}=0\eqn\abc$$

Thus the action \acta\ can be obtained from gauge-fixing the Schild action
\sac\ with $ \nu =1/4$.
Note that the condition $det g_{\alpha \bar \beta}= e^{- \fff} $ from \fiis\
together with the field equation $V= \sqrt g$
implies that $V$ can be identified with $ e^{-\fff}$, at least in a special
coordinate system.

Acknowledgements

I would like to thank M. Abou Zeid, D. Kutasov,  E. Martinec  and G.
Papadopoulos for useful discussions and G. Gibbons for drawing my attention to
the references [\klein,\hit].

\refout

\bye